\documentclass[%
reprint,
 amsmath,amssymb,
 aps,
 pra,
floatfix,
]{revtex4-2}

\usepackage{graphicx}
\usepackage{bm}
\usepackage{braket}             
\usepackage[english]{babel}
\usepackage{lipsum}

\DeclareMathOperator{\Tr}{Tr}

\begin{document}


\title{Blind witnesses quench quantum interference without transfer of which-path information}  

\author{Craig S. Lent}
\affiliation{%
Department of Physics\\
and Department of Electrical Engineering\\
University of Notre Dame\\
Notre Dame, IN 46556, USA
}%

\date{\today}

\begin{abstract}
We model the fundamental behavior of a two-branch quantum interference device. Quantum interference oscillations are visible in the output as the magnetic flux through the paths is varied. Multiple witness systems are field-coupled to each branch. Each witness state entangles with the device state, but for our {\em blind} witnesses which-path information is not transferred to the quantum state of witnesses--they cannot ``see'' or make a record of which path is traversed.  Yet the presence of these minimal witnesses rapidly quenches quantum interference. Thus, it is not the imprinting of which-path information in the witness states that is essential for decoherence, but simply the entanglement that embeds the device degrees of freedom in the larger Hilbert space that includes the witnesses.  The loss of interference visibility can be understood as the result of phase cancellations from different paths through the larger state space. 
\end{abstract}

\maketitle








\section{Introduction  \label{sec:Introduction}}

Two slit quantum interference is arguably the paradigmatic quantum effect. Maxima and minima analogous to Young's famous double slit optical diffraction illustrate the importance of the superposition of distinct possible dynamic paths. The vanishing of the interference if one measures which path the particle actually takes is crucial to the mystery. As Feynamnn puts it, ``when we look at the electrons, the distribution of them on the screen is different than when we do not look'' \cite{Feynmann1964}. There is now no doubt both that this is true and that it reveals a fundamental feature of the physical law. The recent Bell test experiments have shown that the quantum indeterminacy which resides in a superposition state is a feature of reality and not just one's knowledge of reality \cite{Shalm2015, Giustina2015, Hensen2015, Zeilinger2018}. 
A measurement does not actually reveal a pre-existing fact, but one could say the measurement forces the physical world to make a choice between physically allowed but distinct possible outcomes. We prescind here from the fundamental issue of the ``measurement problem'': determining at what point between the microscopic and  macroscopic domains one of the allowed outcomes appears and, so to speak, gets written onto the transcript of the history of events.

If a measurement yields an unambiguous result for a specific dynamical path, clear which-path information, then there is no interference at all.  But a measured result is subject to  noise and finite precision, so loss of coherence admits of degrees.  The interference pattern can be reduced in visibility without being eliminated entirely. The growing interest in applications of quantum information has raised the importance of understanding in some detail this decoherence process, which by now has a very broad literature.

Buks {\em et al.} observed electron interference in a two-path system in a semiconductor modulated by a perpendicular magnetic field \cite{Heiblum1998}. A which-path detector composed of a quantum dot and quantum point contact was added to one path. Modulation of the visibility of the interference pattern was correlated with the sensitivity of the detector. The work presented here is roughly modeled on their experiment. Interference in C70 matter waves in a vacuum chamber has been observed and was gradually quenched with increasing partial pressure of argon \cite{Arndt2003b}.  In this case there is not an  actual measurement, but decoherence due to a transfer of which-path information from the C70 molecular state to argon atoms which scatter off the molecule \cite{ZeilingerArndt2003}.

Wootters and Zurek \cite{WoottersZurek1979} applied the tools of information theory to analyze the double slit interference problem for photons, making quantitative the central argument of the Bohr-Einstein debate at the Fifth Solvay Conference. Extending this work, Englert derived an inequality  \cite{Englert1996} that connects the distinguishability which characterizes a which-way detector to the interference visibility.  The complication of multi-slit interference has been examined by Qureshi and others \cite{Qureshi2015}. Broader questions about the nature of the detector or ``meter'', its disturbance on the observed system, and the roles of uncertainty and complementarity have all impacted the discussion.  Recently, addressing quantum measurement,  Patekar and Hofmann \cite{PatekarHofmann2019}  emphasized the need to  clearly distinguish the role of system-meter entanglement from the projective measurement of distinguishable meter states. 

It is clear that decoherence does not require a projective measurement.  Because of the success of Zurek's paradigm,  decoherence is now commonly described in terms of the movement of information, using terms such as ``information deposited in the environment'' and ``environment as witness" \cite{Ollivier2005}, or ``information transfer from the system to the environment'' \cite{Hornberger2009}, etc. 
Information transfer is certainly sufficient to cause decoherence, but is it a necessary condition?

Here we construct a simple model for a two-path quantum interference device with  witnesses on each branch. The model includes a Hamiltonian for the entire coupled system and we solve for the unitary dynamics directly using the time-development operator. No additional approximations are employed.  We solve for the motion of a wave-packet traversing the device and examine the interference pattern at the output as a function of an applied magnetic field. The witnesses are constructed so that they {\em could} be used as ``meters'' in the sense of Pateka and Hoffman---ancillary systems whose projective measurement yields which-path information---but they are additionally constrained so they are not actually able to sense or record this information. We therefore call these ancilla {\em blind} witnesses.  

We distinguish three requirements for a which-path measurement scheme: (1) the primary system is dynamically coupled to the witness, (2) the dynamics is such that the relevant which-path information is transferred to the witness degrees of freedom, and (3) a projective measurement of the witness results in classically available which-path information.  We explore the effect of minimal witnesses on the visiblity of quantum interference when requirements (2) and (3) are absent. 

The following section describes the model of the quantum interference device in the absence of witnesses and in Section \ref{sec:Dynamics} we solve for the dynamics of the system. Section \ref{sec:DynamicsWithWitnesses} adds the witnesses and shows the effect they have on the quantum interference. In Section \ref{sec:DynamicsOfWitnesses} we examine the dynamics of the  witnesses themselves to see how their quantum degrees of freedom evolve.  The discussion in Section \ref{sec:Discussion} explains how the strong effect of even blind witnesses in suppressing quantum interference can be understood, and the connection to environmental decoherence. 



\section{Model system \label{sec:ModelSystem}}

\begin{figure}[bt]
\centering
\includegraphics[width=8.6cm]{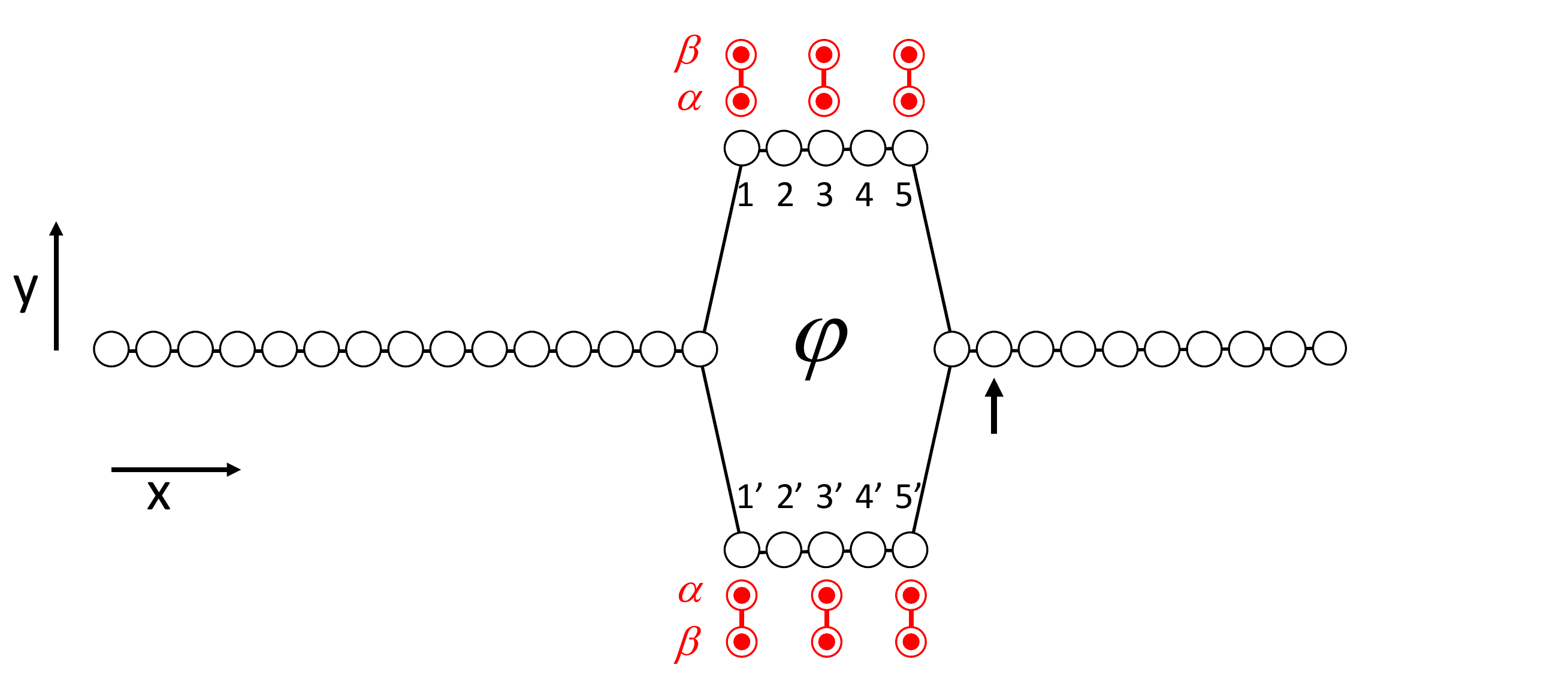}
\caption{Model quantum interference device  geometry.  A tight-binding type site model with near-neighbor tunnel coupling represents the interference device. The input lead on the left is connected to upper and lower branches with the output on the right. A perpendicular magnetic field creates a magnetic flux $\varphi$ through the loop. An input wavepacket is injected from the left and emerges on the right. The arrow on the right indicates the output site $j_{out}$. Each witness consists of two dots, labeled $\alpha$ and $\beta$, which are field-coupled to sites on the upper and lower branches. The presence of an electron at the adjoining device site raises the energy if the $\alpha$ site of the witness is occupied. The geometry illustrated is the case of 6 witnesses at sites $[1, 1', 3, 3', 5, 5']$. Minimal ``blind''  witnesses are constrained to have equal occupancy probabilities for the $\alpha$ and $\beta$ sites as illustrated here by the solid (red) circles.
\label{fig:DotGeometry}
}
\end{figure}

We consider the basic quantum two-path interference device with the geometry shown in Fig.~\ref{fig:DotGeometry}. We model the device using a tight-binding type approach with $N=35$ discrete sites. A line of sites on the left splits into  top and bottom branches and then merges again on the right. (At this point, ignore the double-dot witnesses on the top and bottom branches shown in the figure.)  We  consider an electron  (with charge $-e$) incident from the input lead on the left and examine the output on the right. Let $\Ket{j}$ represent the state with the particle localized on the $j^{th}$ site at position $\vec{r}_j= (x_j, y_j)$. Site 1 on the left edge of the input lead is at the origin of the coordinate system. The  $x$-coordinates of the sites are integer multiples of a distance $a$, though the magnitude of $a$ will finally play no role. The $y$-coordinates are 0 for the input and output leads, and $\pm a/2$ for the top and bottom branches. For clarity in labeling, the vertical and horizontal distant scales in Fig.~\ref{fig:DotGeometry} are different. 
The index $j$ runs from 1 to 15 for the input lead, from 16 to 20 for the top branch,  21 to 25 for the bottom branch, and from 26 to 35 for the output lead. The branch sites are additionally  (redundantly) labeled $1$ through $5$ in the top branch and $1'$ through $5'$ in the bottom branch for convenience in discussing the witnesses below. 

A uniform perpendicular magnetic field $\vec{B}$, described by the vector potential $\vec{A}=-By\hat{x}$, produces different quantum phase shifts for the electron traversing top or bottom branches, creating varying amounts of constructive or destructive interference at the output.  The magnetic flux through the loop formed by the top and bottom branches is $\varphi$. The strength of the applied field can be specified by choosing $\varphi/\varphi_0$, where $\varphi_0 = 2\pi\hbar/e$ is the magnetic flux quantum.

The input lead, top and bottom branches, and output lead constitute the interference device (as distinct from the witnesses) and we can write the device Hamiltonian:
\begin{equation}
    \hat{H}_d=   \sum_{i,j} 
     t_{i,j}  \Ket{i} \Bra{j} + t^*_{i,j} \Ket{j}  \Bra{i}.
     \label{eq:Hamiltonian0}
\end{equation}
\noindent The hopping matrix element $t_{i,j}$, using the Peierls substitution \cite{Peierls1933} to account for the magnetic field, is given by
\begin{equation}
     t_{i,j}  = -\gamma\; e^{-i\frac{e}{\hbar} \int_{\vec{r}_j}^{\vec{r}_i} \vec{A}\cdot d\vec{r}}
     \label{eq:tij}
\end{equation}
\noindent for sites $i$ and $j$ which are connected as indicated by the lines in Fig.~\ref{fig:DotGeometry},  and zero otherwise. At the points where the input and output leads connect to the two branches, the magnitude of the connecting matrix element remains $\gamma$ for simplicity, even though the distance between sites is $a\sqrt{2}$. 

\section{Dynamics without witnesses \label{sec:Dynamics}}

Given an initial state, we find the state at a future time $t$  directly using the unitary time evolution operator.
\begin{equation}
    \Ket{\psi(t)}=e^{-i\frac{\hat{H}_d}{\hbar}t} \Ket{\psi(0)}
    \label{eq:UnitaryHsimple}
\end{equation}
\noindent The natural time scale $\tau$ of the motion depends on the the magnitude of the hopping matrix element $\gamma$.
\begin{equation} 
   \tau\equiv \pi\hbar/\gamma. 
   \label{eq:tau}
\end{equation}

Consider a wave packet initially in the input lead and moving to the right. At time $t=0$ we set the wave function on the input lead to be
\begin{equation}
    \Braket{j|\psi(0)}=A\; e^{-(x_j-x_0)^2/(2w^2)}\; e^{ikx_j}.
    \label{eq:psi0}
\end{equation}
\noindent Here we take $x_0=5a$, $w=2a$,  $ka=\pi/2$, with $A$ chosen for normalization. 

\begin{figure*}[bt]
\centering
\includegraphics[width=12cm]{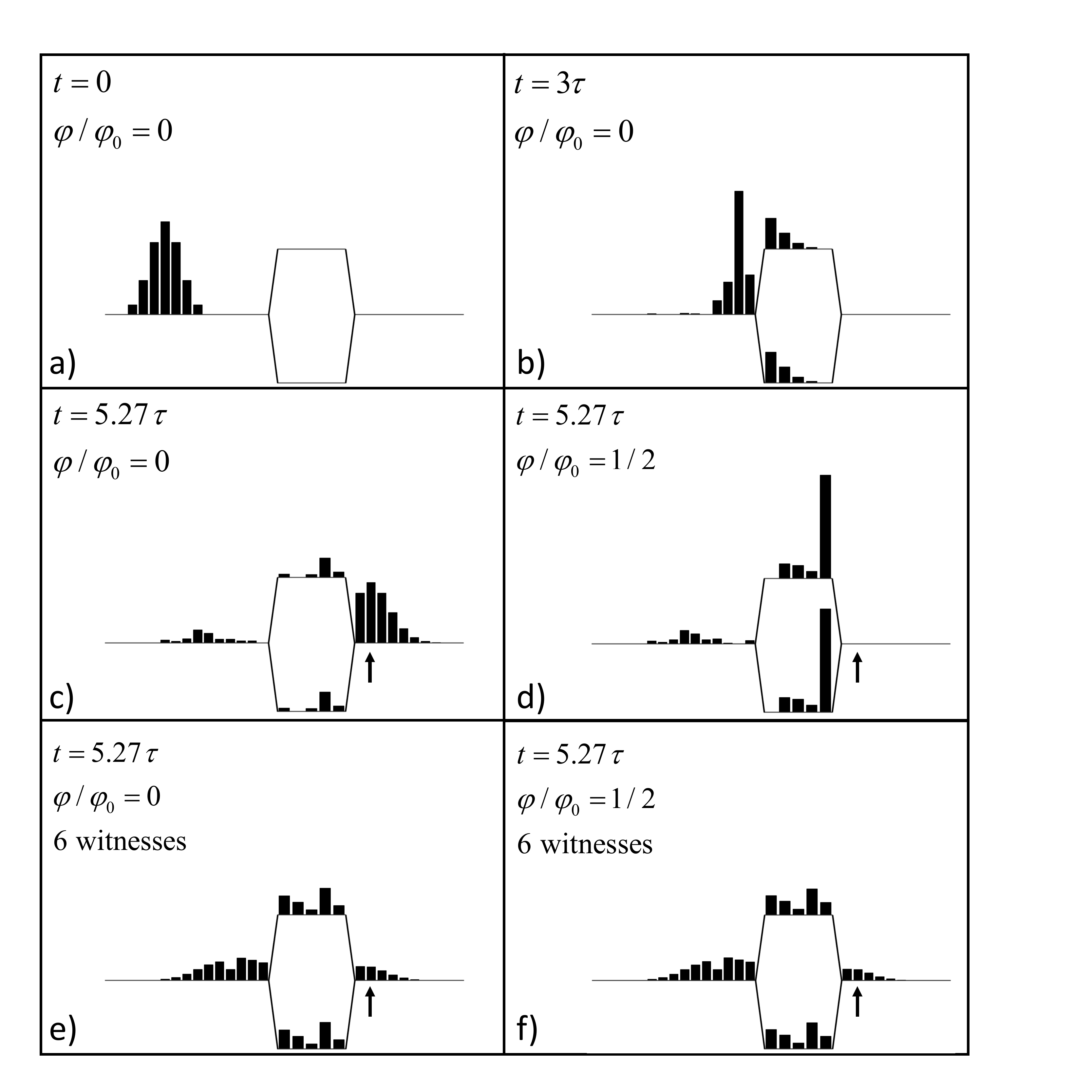}
\caption{Snapshots of the probability distribution. The probability at each site shown in Figure \ref{fig:DotGeometry} is represented by the  height of the solid bar at that site. Times are represented in terms of the characteristic time $\tau=\pi\hbar/\gamma$, where $\gamma$ is the magnitude of the site-to-site hopping matrix element.
(a) The initial state with the incoming wavepacket described by Eq. (\ref{eq:psi0}). The magnetic flux is zero. 
(b) The probability distribution at $t=3\tau$.  Some reflection from the y-branch on the left is evident. 
(c) At $t=T_f \equiv 5.27\tau$, the wavepacket has emerged into the output lead. The arrow indicates the output site $j_{out}$. For zero magnetic flux, the wavefunction from top and bottom branches interfere constructively.
(d) The same situation as (c), but with  magnetic flux $\varphi/\varphi_0=1/2$. The phase accumulated traversing the top and branches is exactly opposite, leading to completely destructive quantum interference and zero output. 
(e) The zero field case analogous to (c), but with six minimal witnesses. The witnesses are coupled in the geometry shown in Fig.~\ref{fig:DotGeometry} with coupling energy $E_{int}/\gamma = 5$. 
(f) The six-witness case with $\varphi/\varphi_0=1/2$ analogous to (d). The presence of the witnesses in (e) and (f) dramatically reduces the coherent quantum interference evident in (c) and (d). 
\label{fig:ProbabilitySnapshots} 
}
\end{figure*}

Consider first the case where $\varphi/\varphi_0 =0$, no applied field.   Figure \ref{fig:ProbabilitySnapshots}(a) shows the initial $t=0$ probability distribution for the wave packet given by (\ref{eq:psi0}). The input lead is long enough  to assure that the  initial state has no amplitude in the two branches. A snapshot of the probability density at $t=3\tau$ is shown in Fig.~\ref{fig:ProbabilitySnapshots}(b). The left y-branch point that connects top and bottom leads causes some scattering back to the input. Such branching must cause reflections as noted in reference  
\cite{ExnerSeba1989}.

Figure \ref{fig:ProbabilitySnapshots}(c) shows the probability distribution at $t=T_f\equiv 5.27 \tau$. This time is chose so that the peak of the wave packet is at the second site from the left in the output lead (index $j=27$). We denote the basis state for this site $\Ket{j_{out}}$ and take the signal output to be the probability for the electron to be found at this site at $T_f$.
\begin{equation}
    P_{out} = \left| \Braket{j_{out}|\psi(T_f)} \right|^2
    \label{eq:Pout}
\end{equation}{}
\noindent The output site is indicated by the vertical arrow in  Figs. \ref{fig:DotGeometry} and \ref{fig:ProbabilitySnapshots}(c--f). The remainder of the output lead is long enough that the effects of reflection from the right end is negligible.  Note that in Fig.~\ref{fig:ProbabilitySnapshots}(c) some  probability density has back-scattered from the right y-branch back into both top and bottom branches. 

When $\varphi/\varphi_0$ is zero or any integer, the result is identical to that shown in Fig.~\ref{fig:ProbabilitySnapshots}(c)---there is constructive interference at the output. When $\varphi/\varphi_0=1/2$, or any odd half integer, interference between the two paths is completely destructive and the output is zero ,  as shown in Fig.~\ref{fig:ProbabilitySnapshots}(d).  

\begin{figure*}[bt!]
\centering
\includegraphics[width=14cm]{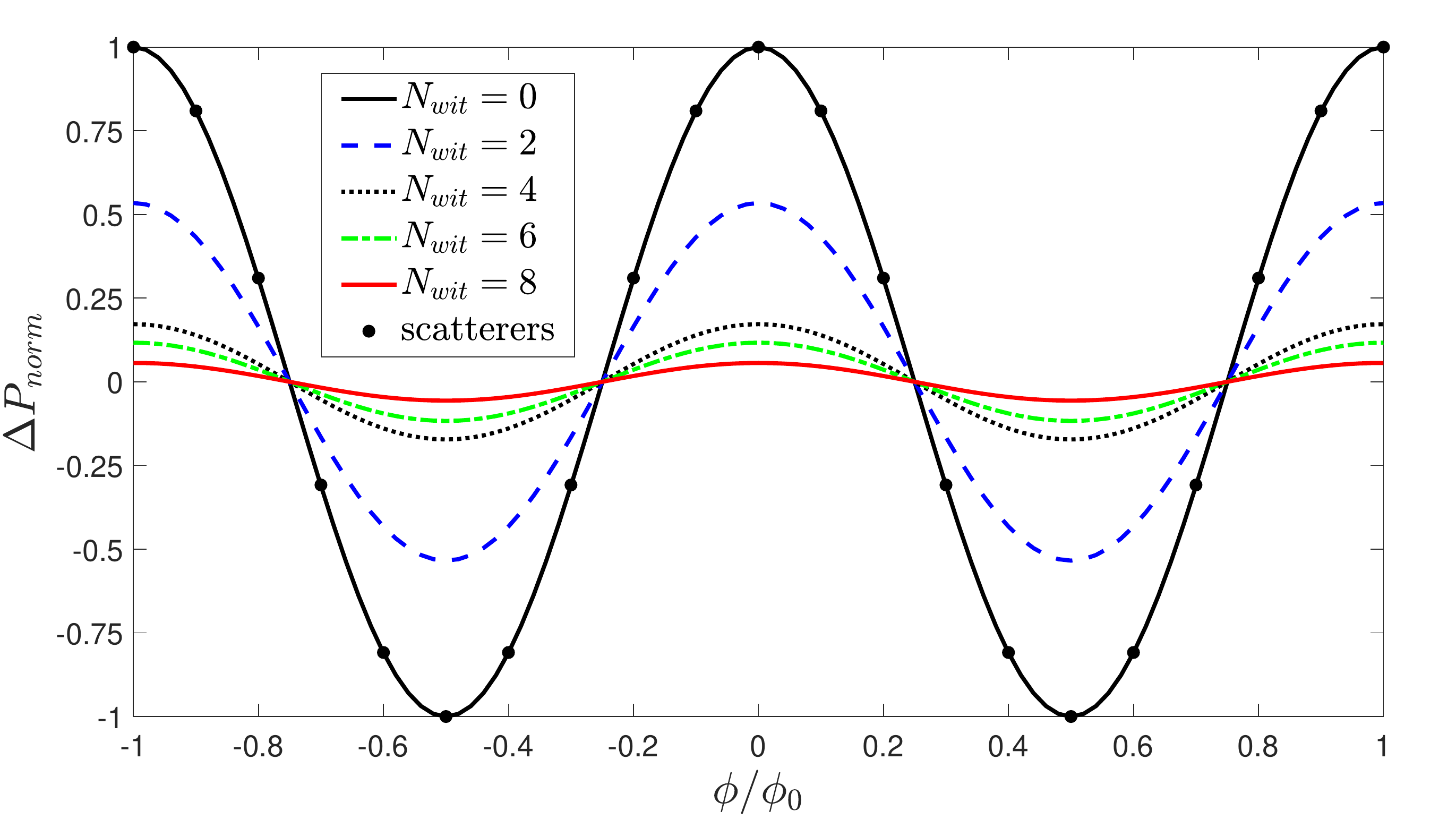}
\caption{Quantum interference in the presence of blind witnesses. The values of the normalized output probability (Eqs. (\ref{eq:Pout}) and (\ref{eq:PnormDef})) are shown as a function of the magnetic flux for different numbers of witnesses. The visibility $\mathcal{V}$ of the normalized interference pattern is half the peak-to-peak value.  The interaction energy between the device and each witness is $E_{int}=5\gamma$.  The solid circles are  for the case when there are no witnesses, but instead fixed scatterers at the $[1,1',3,3',5,5']$ sites. The values fall exactly on the curve for zero witnesses. 
}
\label{fig:InterferencePhi}
\end{figure*}   

The solid (black) line in Fig.~\ref{fig:InterferencePhi} shows the normalized output $\Delta P_{norm}$ as a function of magnetic flux through the loop $\varphi/\varphi_0$.  Let $P_{max}$  and $P_{min}$ be the maximum and minimum value of $P_{out}(\varphi)$. The midpoint value is then $P_{mid}=\left(P_{max}+ P_{min}\right)/2$ and the normalized output is 
\begin{equation}
    \Delta P_{norm} = \frac{P_{out}-P_{mid}}{P_{mid}}
     \label{eq:PnormDef}
\end{equation}

Figure \ref{fig:InterferencePhi} shows that $P_{norm}$ is periodic in $\varphi/\varphi_0$ with maximal interference fringes that extend from -1 to +1. The interference visibility, defined by
\begin{equation}
    \mathcal{V} = \frac{P_{max}-P_{min}}{P_{max}+P_{min}},
     \label{eq:VisibilityDef}
\end{equation}
\noindent is half the peak-to-peak value of $P_{norm}$ and in this case is equal to 1.

\section{Dynamics with Witnesses \label{sec:DynamicsWithWitnesses}}

Our approach here is not to perform a measurement on either one or both paths, but to let the particle in each path interact with $N_{wit}$ witnesses that we describe quantum mechanically as part of the same overall system. We model each witness as a quantum double dot Coulombically coupled to  the main system, as shown schematically in Fig.~\ref{fig:DotGeometry}.   The two basis states of the $m^{th}$ witness are $\Ket{\alpha_m}$ and $\Ket{\beta_m}$ representing the states with the particle localized completely on one or the other dot. The $\alpha$ state always denotes the dot closest to the device and which is field-coupled to the nearest device site.  The quantum state of the $m^{th}$ witness can be written as a superposition of these basis states:
\begin{equation}
    \Ket{\phi_w^{(m)}(t)}= a_m(t) \Ket{\alpha_m} + b_m(t) \Ket{\beta_m}.
    \label{eq:PhiWitness}
\end{equation}

We choose the zero of energy so the onsite energy for each dot is zero, and therefore write the Hamiltonian for each witness in isolation as simply
\begin{equation}
    \hat{H}^{(m)}_w = -\gamma_w \left( \Ket{\alpha_m} \Bra{\beta_m}  
                      + \Ket{\beta_m} \Bra{\alpha_m} \right).
                      \label{eq:Hwitness}
\end{equation}
\noindent  In the absence of interactions between the device and the witnesses, the Hamiltonian for the combined system can be written 
\begin{equation}
    \hat{H_c}= \hat{H}^{(1)}_w \otimes \hat{H}^{(2)}_w \cdots
                \otimes \hat{H}^{(N_{wit})}_w \otimes \hat{H}_D.
\end{equation}
\noindent The quantum state of the combined system is the direct product of the individual witness states and the device state. 
\begin{equation}
    \Ket{\Psi}=\Ket{\phi_w^{(1)}}  \otimes  \Ket{\phi_w^{(2)}}  \cdots
                \otimes \Ket{\phi_w^{(N_{wit})}} \otimes \Ket{\psi}
                \label{eq:DirectProductPSI}
\end{equation}

The Hamiltonian representing the interaction between a charge on the $\alpha$ dot of the $m^{th}$ witness and a charge on the nearest device site $j$ is
\begin{multline}
        \hat{H}^{(m,j)}_{int} =E_{int}\; \hat{I}^{(1)}_w \otimes \hat{I}^{(2)}_w  \cdots
                  \otimes \left( \Ket{\alpha_m} \Bra{\alpha_m} \right)\\
                  \cdots
                  \otimes \hat{I}^{(N_{wit})}_w \otimes 
                  \left(  \Ket{j}\Bra{j}\right).
                  \label{eq:InteractionH}
\end{multline}
\noindent Here $E_{int}$  is the interaction energy and $\hat{I}^{(m)}_w$ is the identity operator for the $m^{th}$ witness state.  There is no tunneling between device sites and witnesses; the interaction is purely Coulombic. 
Consider, for example,  the witness shown at the left edge of the top branch in Fig.~\ref{fig:DotGeometry} at the position labeled $1$. The interaction couples the first witness ($m=1$) of six witnesses to the nearby device site (in this case the site with index $j=16$).

A simple classical picture of the function of the witness as an electrometer which could determine which-path information is as follows.  Suppose an electron is moving through one branch of the device and is momentarily resident on the $j^{th}$ site. If at the same time the witness charge is on the nearby $\alpha_m$ dot, then there is an increase of $E_{int}$ in the energy of the system because of the Coulomb interaction.   The witness charge would then be pushed off the $\alpha$ dot onto the $\beta$ dot and a measurement of the occupancy of either would reveal which path through the device the electron had taken. 

The complete Hamiltonian is composed of the combined device and witness Hamiltonian (without interactions) and the interaction term for each pair consisting of a witness $m$ and its associated device site $j$. 
\begin{equation}
    \hat{H}=\hat{H_c} + \sum_{pairs\; m,j} \hat{H}^{(m,j)}_{int}
    \label{eq:TotalHamiltonian}
\end{equation}

\noindent The time evolution of  $\Ket{\Psi}$ is calculated directly using
\begin{equation}
    \ket{\Psi(t)}=e^{-i \frac{\hat{H}}{\hbar}t  }\Ket{\Psi(0)}.
    \label{eq:TotalTimeDevelopment}
\end{equation}

To calculate the probability density at each device site $j$, we embed corresponding  projection operator in the larger Hilbert space that describes the whole system.
\begin{equation}
    \hat{P}_j = \hat{I}^{(1)}_w \otimes \hat{I}^{(2)}_w  \cdots
                  \otimes \hat{I}^{(N_{wit})}_w \otimes 
                  \left(  \Ket{j}\Bra{j}\right)
\end{equation}
\noindent 
The probability of finding the particle at device site $j$ is then
\begin{equation}
    P_j(t)=\Braket{\Psi(t)|\hat{P}_j|\Psi(t)}.
\end{equation}

Now to make the witnesses minimal blind witness elements we add the further restriction that the tunneling energy $\gamma_w=0$. This means that the witnesses have no internal dynamics---all the matrix elements of $\hat{H}_w$ in the $\{\Ket{\alpha}, \Ket{\beta}\}$ basis are  zero. The occupancy of the $\alpha$ and $\beta$ sites cannot change, so $|a_m|^2=|b_m|^2=1/2$ in Eq.~(\ref{eq:PhiWitness}). This does not mean that the witnesses have no quantum dynamics; they entangle with the device system through (\ref{eq:InteractionH}) and there are quantum mechanical degrees of freedom associated with the phases of $a_m$ and $b_m$. The entangled system, of course, cannot in general be factorized into quantum states of the witnesses and the quantum states of the device.

The Hamiltonian of the system is determined by the number of witnesses $N_{wit}$, the positions where they are attached to the device, the hopping energy $\gamma$ (which sets the time scale $\tau$ of the motion), the relative strength of the interaction $H_{int}/\gamma$, and the magnetic flux through the loop $\varphi/\varphi_0$.

We solve the time evolution of the device with the incoming wave packet given by the initial device state (\ref{eq:psi0}), as in Sec. \ref{sec:Dynamics}, but now in the presence of witnesses symmetrically attached to top and bottom branches.  We take the initial state of each witness to be the symmetric ground state:
\begin{equation}
    \Ket{\phi_w^{(m)}(0)}= \left( \Ket{\alpha_m} +  \Ket{\beta_m}\right)/\sqrt{2}.
    \label{eq:WitnessPhiInit}
\end{equation}

Figures \ref{fig:ProbabilitySnapshots}(e) and (f) show snapshots of the probability for the same time $T_f$ as are shown in \ref{fig:ProbabilitySnapshots}(c) and (d), but now in the presence of 6 witnesses in the top and bottom branches. The witnesses are at positions $[1, 1', 3, 3', 5, 5']$, as shown in Fig.~\ref{fig:DotGeometry}, and $E_{int}= 5\gamma$. Figure~\ref{fig:ProbabilitySnapshots}(e) shows the snapshot when $\varphi/\varphi_0=0$ and Fig.~\ref{fig:ProbabilitySnapshots}(f) shows $\varphi/\varphi_0=1/2$. 

In contrast with Figs. \ref{fig:ProbabilitySnapshots}(c, d), the  probabilities shown in Figs. \ref{fig:ProbabilitySnapshots}(e) and (f) are very similar. The presence of minimal witnesses has quenched both constructive interference for $\varphi/\varphi_0=0$ and destructive interference for $\varphi/\varphi_0=1/2$. The probability densities are not actually identical in the two cases, just very similar. 

Each witness causes reflection of the wave packet in the device. 
The probability density in the branches is the result of multiple reflections from both the y-branches on either end and from the witnesses.  For both Figs. \ref{fig:ProbabilitySnapshots}(e) and (f) however, the probability distribution in the top and branches of the device are identical.

Figure \ref{fig:InterferencePhi} shows the normalized output probability for magnetic flux $\varphi/\varphi_0 \in [-1, 1]$ for different number of witnesses. For the case of 2 witnesses, they are at the  $[3, 3']$ positions shown in Fig. \ref{fig:DotGeometry}. For the 4 witness case, the witnesses are positioned at $[1, 1', 5, 5']$, for the 6 witness case, the witnesses are positioned at $[1, 1', 3, 3', 5, 5']$,
and for 8 witnesses the positions are  $[1, 1', 2, 2', 4, 4', 5, 5']$. As the number of witnesses increases, the interference visibility $\mathcal{V}$ is quenched. It is remarkable that with only 8 minimal witnesses, this fundamental quantum interference is so strongly reduced. Insofar as the witnesses can be thought of as representing the effect the environment, a very small and minimal environment is very effective at suppressing interference. It should be emphasized that  the whole system remains coherent, though the entropy of the device and of each witness increases, as we will see in the next section.

\begin{figure}[bt!]
\centering
\includegraphics[width=8.6cm]{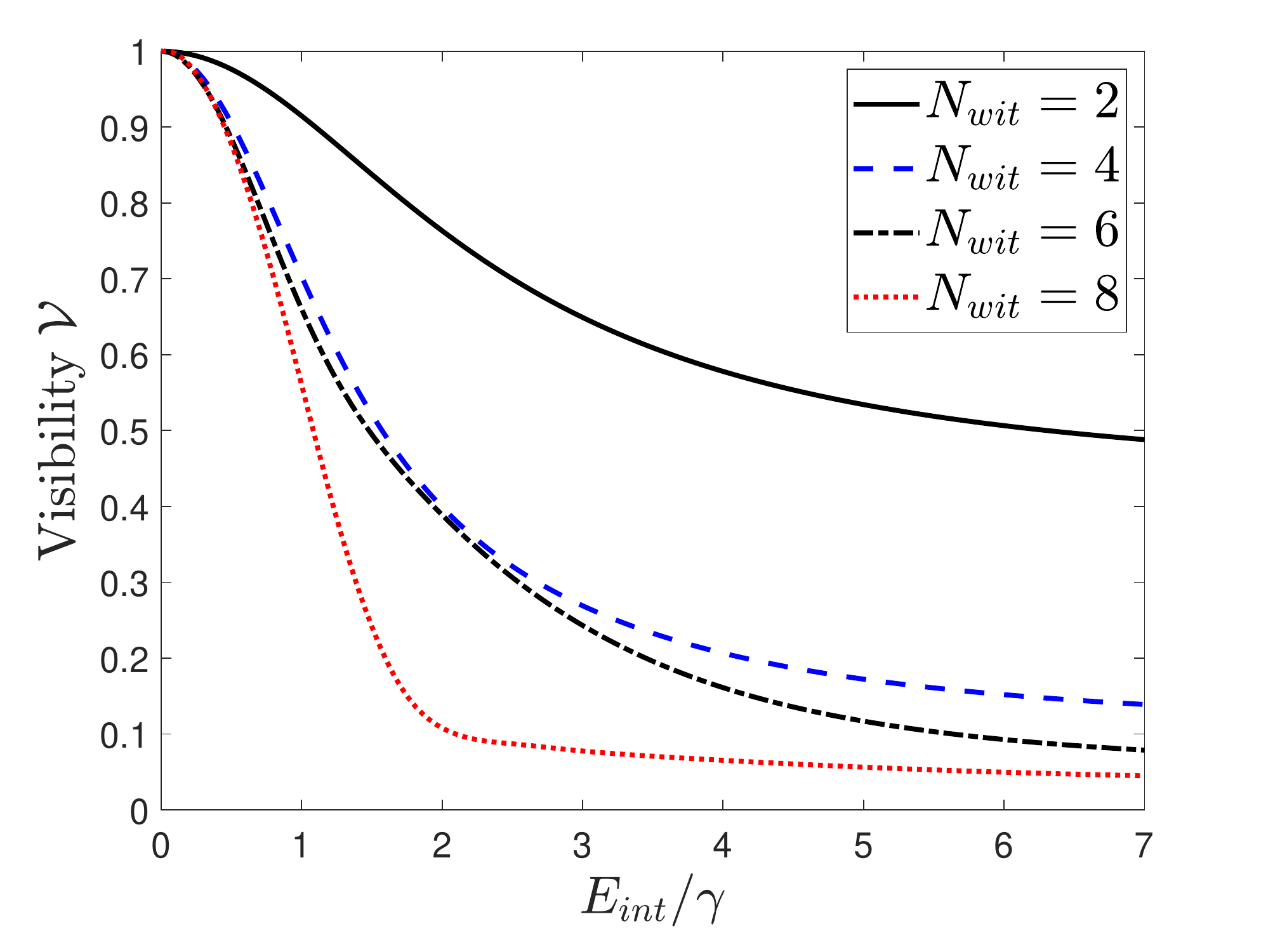}
\caption{Visibility of quantum interference. The visibility given by Eq. (\ref{eq:VisibilityDef}) is plotted as a function of the interaction strength between the device and the witnesses $E_{int}$ for different numbers of minimal witnesses. 
}
\label{fig:Visibility}
\end{figure}   

Figure \ref{fig:Visibility} shows the visibility of the interference as a function of $E_{int}/\gamma$, the scaled interaction energy between the device and the witnesses. The figure shows the result for $2, 4, 6$, and $8$ witnesses, always symmetrically placed in the top and bottom branches. Increasing the strength of the interaction decreases the visibility, though not without limit. For the six-witness case, for example, the visibility at $E_{int}/\gamma=5$ is 11.7\%. For $E_{int}/\gamma=50$ visibility decreases  to 4.8\%, and is essentially the same for $E_{int}/\gamma=500$.  

Because the witnesses necessarily cause scattering in the branches, one might wonder if the scattering by itself is the source of the observed decoherence. The solid dots in  Fig.~\ref{fig:InterferencePhi} show the output when the witnesses are removed and replaced with potential scatterers. We add a scattering term $\hat{H}_s$ to the device Hamiltonian of Eq.~(\ref{eq:Hamiltonian0}):

\begin{equation}
    \hat{H}=\hat{H}_d + \sum_{k_s} \Ket{k_s} V_s  \Bra{k_s},
    \label{eq:HamiltonianScattering}
\end{equation}

\noindent where $V_s=5 \gamma$, and the index $k_s$ runs over the  
$[1, 1', 3, 3', 5, 5']$ sites.  

The resulting interference pattern has visibility $\mathcal{V}=1$; the interference pattern is the same as that of the $N_{wit}=0$ case.  Scattering alone is not causing the quenching of quantum interference---it takes the presence of witnesses.


\section{Dynamics of  Witnesses \label{sec:DynamicsOfWitnesses}}

The classical degrees of freedom of each witness are frozen because $\gamma_w=0$. Dot occupancy cannot change, but the quantum degrees of freedom are affected by the passage of the device electron. For each witness $m$, we define the coherence operators $\hat{\lambda}^{(m)}_x $ and $\hat{\lambda}^{(m)}_y$ of the two-state witness system in the full system by embedding the Pauli operators for the $m^{th}$ witness, $\hat{\sigma}_x^{(m)}$ and $\hat{\sigma}_y^{(m)}$, in the larger Hilbert space.

\begin{eqnarray}
    \hat{\lambda}^{(m)}_x &=& \hat{I}^{(1)}_w \otimes \hat{I}^{(2)}_w  \cdots
                  \otimes \hat{\sigma}_x^{(m)} 
                  \cdots
                  \otimes \hat{I}^{(N_{wit})}_w \otimes \hat{I}_D \nonumber \\
    \hat{\lambda}^{(m)}_y &=& \hat{I}^{(1)}_w \otimes \hat{I}^{(2)}_w  \cdots
                  \otimes \hat{\sigma}_y^{(m)} 
                  \cdots
                  \otimes \hat{I}^{(N_{wit})}_w \otimes \hat{I}_D \nonumber\\     
      \hat{\lambda}^{(m)}_z &=& \hat{I}^{(1)}_w \otimes \hat{I}^{(2)}_w  \cdots
                  \otimes \hat{\sigma}_z^{(m)} 
                  \cdots
                  \otimes \hat{I}^{(N_{wit})}_w \otimes \hat{I}_D   
\end{eqnarray}

We can then calculate the components of the coherence (Bloch) vector  $\vec{\lambda}$ for each witness $m$. 
\begin{equation}
    \lambda^{(m)}_x = \Braket{\Psi|\hat{\lambda}^{m}_x| \Psi}, \quad 
    \lambda^{(m)}_y = \Braket{\Psi|\hat{\lambda}^{m}_y| \Psi}
    \label{eq:LambdaXY}
\end{equation}
\noindent  For the initial witness states given by (\ref{eq:WitnessPhiInit}), 
$\braket{ \hat{\lambda}^{(m)}_z}=0$, and the lack of tunneling between witness sites assures that it will remain zero at all subsequent times. 

From $(\lambda_x^{(m)}, \lambda_y^{(y)})$, we can construct the  $2\times 2$ reduced density matrix for each witness. 
\begin{equation}
    \bm{\rho}^{(m)} = \frac{1}{2}\left(\mathbf{1} + 
    \lambda^{(m)}_x \bm{\sigma}_x + 
    \lambda^{(m)}_y \bm{\sigma}_y\right)
    \label{eq:DensityMatrix}
\end{equation}
\noindent The density matrix represents the best possible local description of the quantum state of the witness.

The local state of each witness can be completely characterized by the $x$ and $y$ components of the coherence vector.  It is helpful to recast the information contained in these two real parameters in another form. We define the coherence angle $\theta_m$ as the  angle the coherence vector of the $m^{th}$ witness makes with the x-axis.
\begin{equation}
    \theta_m = \arctan{(\lambda^{(m)}_y/\lambda^{(m)}_x)}.
    \label{eq:PhaseAnglem}
\end{equation}

\noindent From the density matrix  (\ref{eq:DensityMatrix}) we can also calculate the von Neumann entropy (in bits) for the $m^{th}$ witness. 
\begin{equation}
    S_m = -\Tr\left( \bm{\rho}^{(m)} \log_2( \bm{\rho}^{(m)}) \right)
    \label{eq:vonNeumannEntropym}
\end{equation}

\noindent This entropy is directly related to the length of $\vec{\lambda}^{(m)}$. 
The von Neumann entropy $S_m$ represents the number of bits of missing local information about the quantum state of  witness $m$ due to its entanglement with the device system, and through that, to other witnesses \cite{LentPRE2019}. 

\begin{figure}[bt!]
\centering
\includegraphics[width=8.6cm]{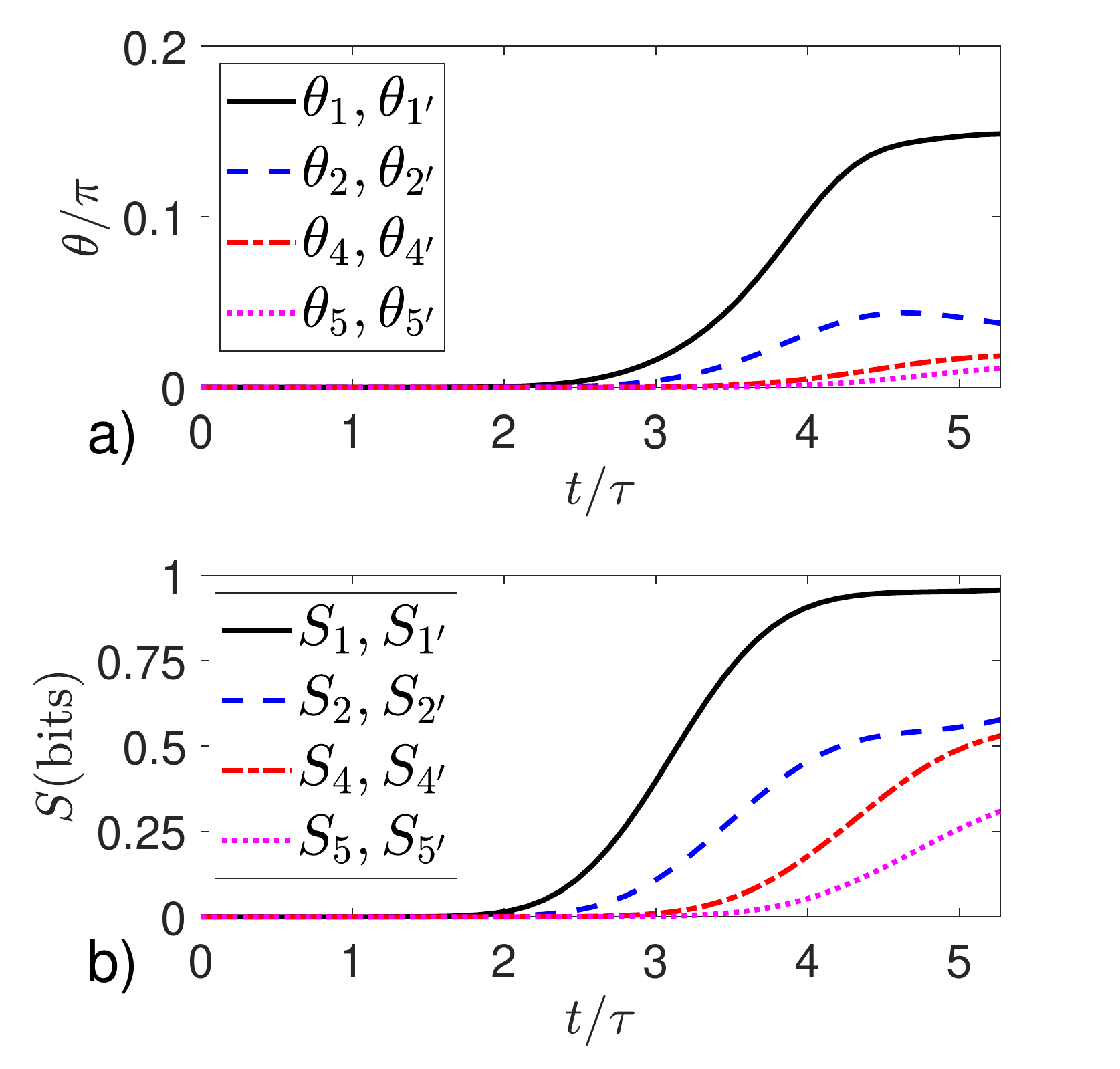}
\caption{Phase angle and entropy of witnesses. The quantum state of each witness is characterized by its phase angle $\theta$  and its von Neumann entropy $S$. The magnetic flux is $\varphi/\varphi_0=1/2$.
(a) The phase angle as a function of time up to $T_f$ for 8 witnesses. Each curve is labeled with the witness position shown in Fig.~\ref{fig:DotGeometry}.  
(b) The corresponding witness von Neumann entropy as a function of time.  Importantly, the quantum state of symmetrically placed witnesses ({\em e.g.}, $1$ and  $1'$) are exactly identical and so do not contain information about which path was taken by the device electron.
}
\label{fig:EntropyAndAngle}
\end{figure}

Figure \ref{fig:EntropyAndAngle}(a) shows the time development of the coherence angles for the device with 8 witnesses, $E_{int}/\gamma=5$, and $\varphi/\varphi_0=1/2$.  
The curves are labeled with the positions of the witnesses shown in Fig.~\ref{fig:DotGeometry}. The dynamics is calculated from equation (\ref{eq:TotalTimeDevelopment}) which yields the global (device plus witnesses) system state $\Ket{\Psi(t)}$ and localized to particular witnesses through (\ref{eq:LambdaXY}). Figure \ref{fig:EntropyAndAngle}(b) shows the entropy $S_m$ for each witness as a function of time.  As the wave packet passes by each witness, the coherence angle shifts slightly and the entropy increases due to entanglement with the device. Corresponding plots for the $\varphi/\varphi_0= 0$ case differ only slightly. 
The entropy of the device itself at $t=T_f$ is approximately 2.5 bits. One could say that the information missing from each subsystem (device and witnesses) is now in the global quantum correlations (mutual information) between them, but that is merely to restate that the information is missing from the subsystems and that $\Ket{\Psi(t)}$ remains pure with $S=0$. 

The time scale shown in Fig.~\ref{fig:EntropyAndAngle} goes to $T_f$, when the peak of the packet enters the output lead and we consider the interference ``experiment'' complete. If the time is extended beyond that, the wave packet bounces back and forth from the ends of the device (beyond the model's primary intent) and the entropy of each witness approaches its maximal value of 1 bit. The extended time for interaction removes nearly all local quantum information from the witnesses and the best local description becomes a purely classical mixture of the two dot occupancies each with probability $1/2$. Figure \ref{fig:EntropyAndAngle}(b)  shows the beginning of that process. The entropy of the device itself also increases to about 3.9 bits in the long run. The maximum possible entropy for the device  would be $S=\log_2(N)\approx 5.13$ bits.  

The most important feature to note in Fig.~\ref{fig:EntropyAndAngle} is that the symmetrically placed witnesses ({\em e.g.}, those at positions $1$ and $1'$ or $2$ and $2'$) are always in exactly identical states. They have the same values of $S_m$ and $\theta_m$ (or equivalently $\lambda_x^{(m)}$ and $\lambda_y^{(y)}$) at all times. This is true at for $\varphi =0$ as well. The witness states bear no asymmetric imprint representing which-path information.  The output interference pattern is destroyed even though the witnesses were unable to ``look'' at which path was taken.  Just the fact of entanglement between witnesses and device is sufficient to destroy the coherent oscillations, as Figs. \ref{fig:InterferencePhi} and \ref{fig:Visibility} demonstrate.


\section{Discussion \label{sec:Discussion} }

The three criteria for which-path measurement we have described in the introduction are: entanglement between the primary system and the  witness,  transfer of which-path information to the witness dynamical degrees of freedom, and measurement of the witness state. If all those things happen, decoherence will certainly result. 

In the system studied here, no measurement of path information is made. The witnesses entangle with the device wavefunction, but there is no projective collapse of the witness state. The dynamics are calculated with the full Hamiltonian of the device plus witnesses using the unitary, reversible, time evolution of equation (\ref{eq:TotalTimeDevelopment}) and the global state remains pure throughout. 

More importantly for our purpose here,  there is no copy of which-path information transferred to the witnesses. This is why we term them minimal witnesses. For both the maxima and minima of Fig.~\ref{fig:InterferencePhi}, which together determine the interference visibility, the quantum states of corresponding witnesses in both top and bottom branches are the same, as shown in Fig.~\ref{fig:EntropyAndAngle}. From the quantum state of the witnesses, represented by their density matrices $\bm{\rho}_m$, one cannot determine which path through the device the electron took. Yet the interference is quenched. The conclusion can only be that neither projective measurement nor which-path information transfer is necessary for decoherence. What is essential is simply entanglement between the device and the witnesses with the resultant expansion of the size of the relevant Hilbert space. Multiple witnesses further quench the interference, reducing the visibility.

Randomness can also cause loss of coherence. If, for example, the environment consists of multiple elements whose interaction energy with the target system has statistical spread, then the phase relationship of different components may average out. This has been shown in a spin system by Cucchietti et al. \cite{Zurek2005} and we have seen similar effects in a double-dot system with precisely the same minimal witnesses as are used here \cite{Blair2013}. No doubt many physical environments  have exactly this character. Just as optical interference effects are masked by the presence of many frequencies, so too many quantum interference effects, though still present in a fundamental sense, are not apparent because of this random averaging. 

But randomness, by design, plays no role in the calculation described here. The witnesses are geometrically regular and the interaction strength between each witness and the device is the same. The initial states of the witnesses are not random. Choosing random values for the initial $\theta_m$, the angle of the coherence vector around the equator of the Bloch sphere, has no effect on the interference or visibilities shown in Figs. \ref{fig:InterferencePhi} and \ref{fig:Visibility}. The only change in the results presented here would be to rigidly offset the $\theta_m(t)$ curves shown in Fig.~\ref{fig:EntropyAndAngle}a by the initial (randomly chosen) value. 

One might consider that the loss of coherence we see could be due to the interaction between the multiple spatial wavelengths (momenta) present in the incoming wavepacket and the reflections caused by the witnesses and y-branches. The finite distribution  of wavelengths could be thought to average out the interference. But if that were the case, we would expect to see interference similarly reduced when witnesses are replaced by static potential scatterers. As discussed above and shown in Fig.~\ref{fig:InterferencePhi}  (points on the solid $N_{wit}=0$ line), no visibility reduction is caused by the presence of  scatterers in the branches. 


The crucial effect of witnesses is to expand the relevant Hilbert space to include both device and witness degrees of freedom. In this model, the device electron is represented in an $N=35$ dimensional space. For the 6-witness case illustrated in Figs. \ref{fig:DotGeometry} and \ref{fig:ProbabilitySnapshots}(e,f), for example, the space is expanded to a dimension of $2^6\times 35$, including the device states $\{  \ket{j}\}$, and the localized states $\{\ket{\alpha_m}\}$ and $\{\ket{\beta_m}\}$ for each witness $m$. 

\begin{figure}[tb]
\centering
\includegraphics[width=8.6cm]{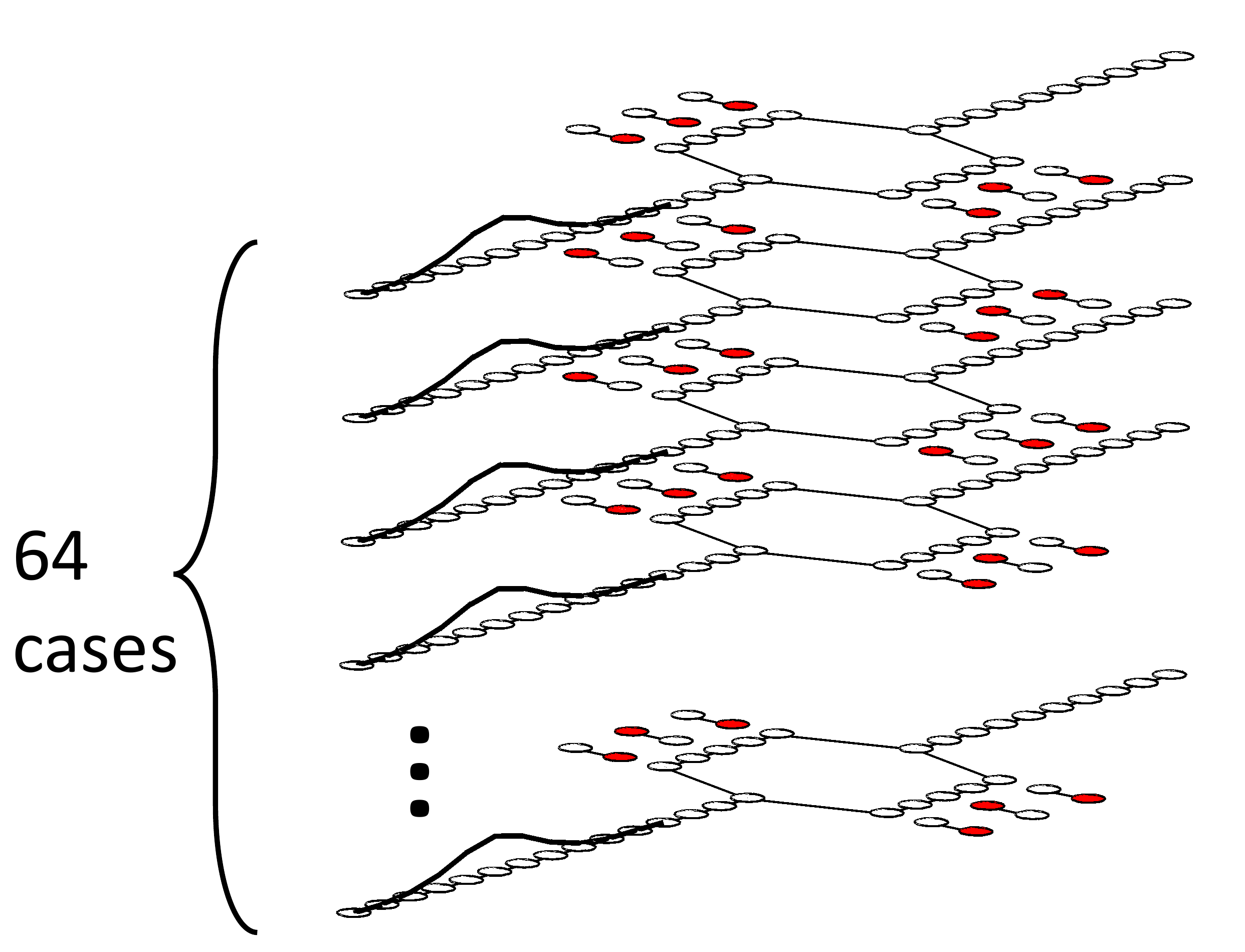}
\caption{Schematic of the enlarged Hilbert space. For the  six-witness case shown, the relevant Hilbert space includes the states of the interference device and all $64=2^6$ possible localized states of the witnesses. These are shown here factored into a stack of replicas of the device with different witness configurations. The Hamiltonian does not couple the two localized witness states $\Ket{\alpha_m}$ and $\Ket{\beta_m}$, so  unitary time evolution of each layer in the stack is independent. The probability amplitude of the initial incoming wavepacket is equally distributed among the layers. For each layer the particular configuration of localized witness states produces a different pattern of reflections in the device. The amplitudes for each site add coherently to form the overall wavefunction. The complex cancellations of phase between these different possible paths through the Hilbert space results in the quenching of coherence in the interference pattern seen in Fig.~\ref{fig:InterferencePhi}. This phase cancellation, rather than the transfer of which-path information from the device to the witnesses, is the essential cause of decoherence. 
}
\label{fig:InterferenceStack}
\end{figure}   

How does expanding the Hilbert space produce the quenching of interference? For the 6-witness case, one can visualize the possible paths through this space as a stack of 64 layers with replicas of the device states with input, output, and two branches, as shown schematically in Fig.~\ref{fig:InterferenceStack}. Each replica has one of the configurations of the six witnesses in different fully polarized states with either $\ket{\phi^{(m)}_w}=\ket{\alpha_m}$ or $\ket{\phi^{(m)}_w}=\ket{\beta_m}$. The direct product in Eq.~(\ref{eq:DirectProductPSI}) generates all the $2^{N_{wit}}$ possible combinations. Because the Hamiltonian does not connect different witness states ($\gamma_w=0$ ), each layer evolves independently. The initial  wavepacket is distributed equally among the layers in the input branch at $t=0$. Under unitary time evolution, each layer has a different combination of reflections from the specific configuration of witnesses in that layer, with various amplitudes and complex phases as a result.  The probability at the output, or at any device or witness site, is obtained by summing the amplitudes from all the layers and taking the absolute square of the result. The  partial cancellations from different  phase factors in each layer weakens the constructive or destructive interference at the output. Note that the net probability density on each witness site is always $1/2$.  There is nothing random in this process, but the increased number and variety of paths through the Hilbert space results in the degradation of the coherent interference pattern at the output. The reason increasing the number witnesses is so effective in suppressing the interference visibility is that the number of paths increases exponentially with the number of witnesses.

This analysis illuminates and supports Zurek's decoherence and einselection paradigm \cite{Zurek2003}. Elsewhere we have seen the clear emergence of pointer states in an environment of randomly positioned and oriented minimal double-dots, just as are used here, and the resultant quenching of Rabi oscillations, another quintessential quantum effect \cite{Blair2013}. We have similarly seen that an environment of minimal double-dot witnesses are sufficient to produce the loss  of two-particle entanglement in a system undergoing unitary time evolution \cite{LentBlairEntanglement2018}. For many, probably most, environments, individual elements may indeed receive an imprint from the system sufficient to count as an information transfer from  system to environment, and the multiplicity of those copies favors pointer states.   What we see clearly from the present calculation is that the creation of copies of system state information in the state of the environment is not {\em necessary} for decoherence to occur. The essential feature is simply entanglement, or equivalently, the system becoming embedded in the dynamics of the much larger Hilbert space that includes all the witnesses and results in multiple phase-canceling paths through that space.   

The present calculation is, of course, limited to the dynamics of the specific interference device and witness system described, though we would argue this is a paradigmatic system. But it has the power of a counterexample to the notion that it is the transfer of which-path information, either to the environment or to a measurement device, that destroys quantum interference. What is shown here is that entanglement, and the consequent multiplicity of system paths through the larger state space, even with no imprinting of which-path information on the witnesses, is sufficient.




%

\end{document}